\documentclass[preprint]{aastex}
\usepackage{emulateapj5}
\def\stacksymbols #1#2#3#4{\def\theguybelow{#2}
	\def\verticalposition{\lower#3pt}
	\def\spacingwithinsymbol{\baselineskip0pt\lineskip#4pt}
	\mathrel{\mathpalette\intermediary#1}}
\def\intermediary #1#2{\verticalposition\vbox{\spacingwithinsymbol
	\everycr={}\tabskip0pt
	\halign{$\mathsurround0pt#1\hfil##\hfil$\crcr#2\crcr
		\theguybelow\crcr}}}
\def\lta{\stacksymbols{<}{\sim}{2.5}{.2}}
\def\gta{\stacksymbols{>}{\sim}{3}{.5}}

\begin{document}

\title{CREATION OF X-RAY HOLES WITH COOL RIMS IN COOLING FLOWS}

\author{Fabrizio Brighenti$^{1,2}$ \& William G. Mathews$^1$}

\affil{$^1$University of California Observatories/Lick Observatory,
Board of Studies in Astronomy and Astrophysics,
University of California, Santa Cruz, CA 95064\\
mathews@lick.ucsc.edu}

\affil{$^2$Dipartimento di Astronomia,
Universit\`a di Bologna,
via Ranzani 1,
Bologna 40127, Italy\\
brighenti@bo.astro.it}

%\vskip 2.in
%\noindent
%Received:

%\noindent
%PROOFS TO BE SENT TO:

%\noindent
%Lick Observatory

%\noindent
%Santa Cruz, CA 95064

%\noindent
%$^1$UCO/Lick Observatory Bulletin No.

%\vskip3.in
%\noindent
%Short Title: X-ray Holes in Cooling Flows 
%\clearpage
\vskip .2in

\begin{abstract}
The density irregularities and holes visible in many 
{\it Chandra} X-ray images of cluster and galactic cooling flows 
can be produced by symmetrically 
heated gas near the central galactic 
black hole. 
As the heated gas rises away from the galactic center, 
a relatively small number of 
large plumes and bubbles are formed in qualitative 
agreement with the observed features.
The expanding centrally heated gas 
drives a shock into the surrounding gas, 
displacing it radially.
Both computational and analytic results show that 
the ambient gas near the bubble 
is cooled by expansion, 
accounting for the cool rims commonly observed 
around X-ray holes in cooling flows.
\end{abstract}

\keywords{galaxies: elliptical and lenticular, CD -- 
galaxies: active -- 
cooling flows --
X-rays: galaxies -- 
galaxies: clusters: general -- 
X-rays: galaxies: clusters}

%%%%%%%%\clearpage

\section{Introduction}

High resolution X-ray images of 
``cooling flows'' in 
elliptical galaxies taken with the {\it Chandra} 
observatory indicate that the central hot gas in these systems 
is not smoothly distributed, 
but is cavitated  on scales comparable 
to the radio emission
(M87: B\"ohringer et al. 1995; Fabian et al 2001;
Wilson et al 2001;
M84/NGC 4374: Finoguenov \& Jones 2001:
NGC 4636: Jones et al. 2001; Loewenstein et al. 2001;
David et al. 2002;
NGC 507: Forman et al. 2001;
NGC 5044: Buote et al. 2002).
In view of their short dynamical times, $10^7 - 10^8$ yrs, 
these disturbances, and by inference also the 
radio sources, must be highly transient. 
X-ray holes are also seen 
in galaxy clusters on scales of 
$\lta 50$ kpc, often approximately coincident 
with lobes of extended radio emission 
(e.g. 
Hydra A/3C295: McNamara et al. 2000; David et al. 2001;
Allen et al 2001; 
Perseus/NGC 1275/3C84: Churazov et al. 2000; 
B\"ohringer et al. 1993; Fabian et al 2000;
A2052: Blanton et al 2001; A4059: Heinz et al 2002).
The Perseus cluster in particular contains a multitude of 
X-ray holes 
located at random azimuthal orientations.
The X-ray holes are regions in which the gas is
either heated or displaced by a non-thermal plasma of comparable
energy density.
In either case the low-density holes must be buoyant
(e.g. Churazov et al. 2001).

One of the most remarkable features of these holes is that 
the X-ray gas around the rims is often {\it colder} than 
the average nearby ambient gas 
(e.g. Fabian et al. 2000, 2001; Fabian 2002). 
Low temperature rims
have been held as evidence that the surrounding gas
has not been (recently) strongly shocked, 
although shocks must accompany
even a subsonic expansion of the hot gas.
These low entropy rims cannot be understood 
as local gas that was shocked and subsequently lost 
entropy by radiation
(Nulsen et al 2002; Soker, Blanton \& Sarazin 2002), 
but may instead be low-entropy gas that has somehow been 
raised from the center of the flow as these 
authors suggest.

The random location of older, more distant bubbles in Perseus 
and the irregular X-ray images within the central few kpcs 
of elliptical galaxies. 
are apparently inconsistent with 
non-thermal jets having fixed orientations defined by the 
spin axis of massive black holes. 
Conversely, not every strong radio lobe corresponds to an X-ray 
hole, as in Perseus/NGC1275/3C84.

We show here that both the random orientation of X-ray holes and 
their cold rims can be understood if low entropy hot gas 
near the center of a cooling flow is heated by an 
active nucleus in the central elliptical galaxy.

\section{Previous Studies of X-ray Holes}

Several recent theoretical studies of X-ray bubbles 
have examined 
the consequences of introducing heated gas 
in some localized
region away from the center of the cooling flow. 
The 2D hot buoyant bubbles by 
Churazov et al. (2001), for example, 
are accompanied by a column of colder (low entropy) 
gas that moves radially upward near the center of the bubble.
These bubbles float upward in the cooling flow atmosphere
and come to rest at some large radius where the ambient entropy 
is equal to that within the bubble. 
According to Churazov et al, 
rising bubbles capture cold gas and lift it to large 
distances from the center, although if this gas is not heated 
it would be expected to eventually fall back.
Nevertheless, the heating required to form the bubbles
reduces the global entropy gradient in the cooling flow. 
Quilis et al. (2001) studied the evolution of a nearly axisymmetric 
3D bubble produced by heating gas at some finite radius. 
They noticed that the buoyant bubble is surrounded 
by a shell of slightly colder gas which they attributed 
to cooling expansion as the gas is pushed by the 
bubble toward regions of decreasing ambient pressure.
In these studies 
gas is assumed to be heated at some off-center site, 
but the implication is that 
the ultimate source of energy is located in a central 
active black hole.
Heating by jets that proceed directly from the AGN core  
can also create buoyant regions and upwelling of low entropy 
gas, but only along the jet axis
(Reynolds, Heinz, \& Begelman 2001; 2002). 
Reynolds et al. (2002) also 
note that ``curiously, the [observed] 
cool gas seems to form a shell
around the radio lobes and occupies precisely the location 
where we would expect shock/compressionally heated 
gas to reside''.
Regardless of the heating geometry assumed, 
by jets or ad hoc off-center bubbles, 
the global, long-term rate that hot gas cools by 
radiative losses in cooling flows  
is unlikely to change 
without altering the gas density and temperature profiles 
in disagreement with observations. (Brighenti \& Mathews 2002).

\section{Producing Bubbles with Central Heating}

We begin by constructing a simple 
cooling flow for a massive 
elliptical galaxy. 
The gravitational potential is a superposition of 
a de Vaucouleurs stellar mass profile of total 
mass $M_{*t} = 7.26 \times 10^{11}$ $M_{\odot}$ 
and a Navarro-Frenk-White dark halo 
with virial mass $M_{h,vir} = 4 \times 10^{13}$ $M_{\odot}$
and concentration $c = 10$, 
typical for the galaxy 
group environments in which massive E galaxies 
are thought to form.
The initial cooling flow that we consider is generated entirely
by mass loss from an old stellar population, 
$\alpha_* \rho_* = 4.7 \times 10^{-20}(t/t_n)^{-1.3}$ 
gm s$^{-1}$ where $\rho_*(r)$ is the stellar density 
and $t_n = 13$ Gyrs is the assumed age of the stars. 
Some heating is provided by Type Ia supernovae,
exploding at a rate of $0.06 (t/t_n)^{-1}$ per 100 years 
per $10^{10}L_B$, each releasing $10^{51}$ ergs.
Supernova heating at this level is consistent with the iron 
abundance observed in the hot gas but has a relatively minor
influence on the overall energetics of the hot gas. 
The gas-dynamical equations that we solve are identical to 
those described in Brighenti \& Mathews (2002). 
We use a 2D code modeled after ZEUS 
(Stone \& Norman 1992) with 490 
grid zones in both cylindrical coordinates.
For $R,z < 50$ kpc the zones are uniform with size 
$125 \times 125$ pc.
Outside this region the size increases geometrically toward 
the outer boundary at $260 \times 260$ kpc.

The quiescent cooling flow that results 
after slowly evolving for 13 Gyrs 
has a characteristic temperature of $\sim 1$ keV and 
a gas density profile $\rho \propto r^{-1.2}$ within 
about 10 kpc and slightly steeper beyond.
We assume that gas within $r_h = 1$ kpc of the center 
is heated at a constant rate 
$L_h = 2.5 \times 10^{42}$ erg s$^{-1}$ 
for $t_h = 5 \times 10^7$ yrs. 
The heated gas reaches temperatures far in excess of 
1 keV and has a sound crossing time much less than $t_h$, 
allowing an ultra-hot low-density region to develop 
near the galactic core. 
The hot gas pushes outward, driving a shock into the 
cooling flow gas beyond.
The resulting evolution of the gas density and temperature are 
shown in Figure 1 
at three times $t_1 = 40 \times 10^6$, 
$t_2 = 80 \times 10^6$, and $t_3 = 100 \times 10^6$ yrs. 
The Rayleigh-Taylor (RT) development of a few large plumes filled
with heated gas produces 
irregular X-ray patterns 
that resemble {\it Chandra}
images of NGC 5044 and other E galaxies.
Small wavelength RT unstable features 
are not prominent even though they would be  
resolved by the numerical grid.
The hot cavity expands gently and is already subsonic 
($\lta 500$ km s$^{-1}$) at $t = 1$ Myr.
By time $t_1$ the weakened precursor shock has moved
out to $\sim 20$ kpc, moving almost sonically.

Low temperature, bright rims surrounding the hot plumes/bubbles are
clearly visible in Figure 1.  
The peak density in the rims at $t_1$ is
several times that of the initial unshocked gas at this radius.  
The
minimum entropy $\log(T/n_e^{2/3})$ in the simulation at times $t =
0$, $t_1$, $t_2$ and $t_3$ are 7.15, 7.9, 8.1 and 8.0 respectively.
Low entropy gas in the rims at $t \gta t_1$ was originally at $r \sim
r_h$ as verified with passively advected tracer particles.  
This gas
cools by adiabatic expansion while rising in the hot gas atmosphere
(see also Nulsen et al. 2002; Soker et al. 2002).  
The larger bubble
at time $t_2$ is penetrated by a disorganized S-shaped column of
slowly rising cold gas.  
The upper panel of Figure 2 shows the X-ray
surface brightness of the central cooling flow at time $t_2$   
when the plumes have separated into clearly visible buoyant X-ray
bubbles.

It is important to stress that 
the physical size of the plumes that form is not 
determined by our computational resolution.
Calculations repeated at higher spatial resolution 
(central grid size=50 pc) also produce a few 
plumes and (ultimately) bubbles of about the same size 
as those shown in Figures 1 and 2, which are similar to 
the X-ray patterns observed. 
An examination of the effective gravity 
in the gas rest frame indicates 
that the outer rims of the large bubbles are
RT unstable. 
At time $t_3$ 
the gravitational freefall time across the bubble radius is 
$t_{ff} \approx 0.1 t_3$; coherent hot bubbles persist 
for times $\gg t_{ff}$, even at our highest computational 
resolution.

We performed similar calculations on 
more extended cooling flows 
in galaxy groups and clusters with very similar results. 
For example, the lower panel of Figure 2 illustrates the
X-ray surface brightness at time 
$t = 120 \times 10^6$ years
produced in a rich cluster cooling flow 
($M_{vir} = 10^{15}$ $M_{\odot}$ and 
$T_{vir} \approx 6.9$ keV)
with heating parameters:
$L_h = 10^{45}$
ergs s$^{-1}$, $r_h = 2$ kpc and $t_h = 5 \times 10^7$ yrs.
Very little gas cools by radiation losses  
during the brief duration of our calculations shown in 
Figures 1 and 2  
so the flow is essentially adiabatic.
This is 
expected since a defining attribute of cooling flows 
is that the dynamical time is much less than 
the radiative cooling time. 

\section{Similarity Solution}

Cold rims are a generic feature of flows 
forced to expand 
into a medium of increasing entropy. 
This can be illustrated with self-similar flows 
described by the variable $\eta = r^{-\lambda}t$ 
(Courant \& Friedrichs 1948;
Rogers, 1957;
Parker, 1961;
Chevalier \& Imamura 1983).
The adiabatic equations for spherical flow can be written 
in terms of $\eta$ provided
$u = (r/t)U(\eta)$, $\rho = A_o r^{-\alpha} \Omega(\eta)$
and $P = A_o r^{2-\alpha} t^{-2} \Pi(\eta)$. 
The hot region of centrally heated gas 
can be represented 
by an expanding spherical piston at $\eta = \eta_2$ 
that drives 
an advancing shock at $\eta = \eta_1 < \eta_2$.
The solutions $U(\eta)$, $\Omega(\eta)$ and $\Pi(\eta)$ 
can be found by inward integration from post-shock values 
$U_1$, $\Omega_1$ and $\Pi_1$ which are functions 
of $\lambda$, $\gamma = 5/3$ and the Mach number of the 
shock ${\cal M}$ (e.g. Rogers 1957).
We consider the spherical flow that results for 
pistons uniformly expanding ($\lambda = 1$) into 
an isothermal, fully ionized medium ($T_o = 10^7$ K) 
with decreasing electron density 
$n_e = 0.0526 r^{-\alpha}$ cm$^{-3}$ where
$\alpha = 1.2$ 
approximates the gas density observed in 
massive elliptical galaxies 
(e.g. Brighenti \& Mathews 1997).
Figure 3 illustrates the flow generated by three pistons 
driving shocks at $r_1 = 3$ kpc with ${\cal M} = 1.5$, 2 and 3.
In each case the temperature drops sharply near the piston, 
{\it to values even less than} $T_o$, creating cold rims 
where the density rises accordingly.
Post-shock gas that has just cooled back to $T_o$ in Figure 3, 
located at $\log(r/r_1) = -0.179$, -0.108, and 
-0.078 for ${\cal M} = 1.5$, 2 and 3 respectively,
was originally located at 
$\log(r/r_1) = -0.347$, -0.476 and -0.957, 
much closer to the origin. 
At the piston where $U < 1 $ approaches unity, 
the gas temperature decreases as $\chi \equiv
\gamma \Pi / \Omega \propto (1 - U)^{p}$
where $p = 2 \alpha /3(5 - \alpha) = 0.21$ 
when $\lambda = 1$.
The cold gas adjacent to the piston is a relic of 
low entropy gas initially near the center that 
was shocked then cooled by expansion; 
$d \log \rho / d \log t$
is negative throughout the flowing gas in the 
plotted solutions. 
Similarity flows with lower shock Mach numbers 
have flatter density profiles;  
for ${\cal M} = 1.5$ there is a shallow density 
minimum at $\log(r/r_1) = -0.047$, 
indicating weak post-shock RT instability.
However, the constant piston 
velocity assumed in the similarity flows 
differs from the rapidly decelerating 
heated bubble in the computed galactic scale flow.

\section{Final Remarks and Conclusions}

If central heating is responsible for 
the frequently observed X-ray holes and surface
brightness fluctuations in 
typical {\it Chandra} images, as we claim, then 
such AGN-black hole heating may be a nearly universal 
component of cooling flows.
The physical origin and 
nature of this heating -- e.g. shock waves, 
cosmic rays, etc. -- has not yet been explored but 
these details are
not essential to the formation of plumes and bubbles. 
While the production of X-ray irregularities 
is quite generally insensitive to 
the assumed heating parameters, we have noticed 
several trends when $L_h$, $r_h$ and $t_h$ are varied.
Non-linear plumes mature faster when $r_h$ is reduced.
If $r_h$ is too large, the
cold rims are less pronounced. 
However, the possibility of learning about such 
heating details by 
comparing X-ray images with gasdynamical calculations is 
limited if the mean time between heating episodes 
is sufficiently short that the initial 
pre-heated gas is no longer perfectly quiescent as we have
supposed here.

All well-observed cooling flows emit diffuse optical 
Balmer and forbidden lines visible in the 
brightest central regions.
Enhanced H$\alpha$ + [NII]
emission has been observed near the boundaries
of X-ray bubbles
(e.g. Blanton et al 2001; Trinchieri \& Goudfrooij 2002).
It is tempting to conclude that such gas has cooled from 
the hot phase in the cooler, denser 
gas around the rims 
of the plumes or bubbles.
But direct cooling from the hot phase is not supported
by our detailed hydrodynamic calculations which are 
essentially unchanged if radiative cooling losses are 
neglected.
The radiative cooling time is greater than the 
total time of our simulations.
The cooler gas at $T \sim 10^4$ K responsible 
for the observed optical line emission may simply be 
displaced by the plumes and bubbles along 
with the local hot gas.
However,
on longer timescales than we consider here, 
heating in the central 
regions stimulates cooling by thermal instabilities 
in more distant regions of the flow, 
so the time-averaged global cooling rate is 
essentially unchanged 
(Brighenti \& Mathews 2002).

If intermittent AGN heating is 
reasonably frequent, the entire inner cooling flow
will become convective and turbulent
(Brighenti \& Mathews 2002).
Turbulent motions observed in the diffuse
optical lines 
(e.g. Caon et al. 2000) are too energetic on large scales
to be understood in terms of supernova explosions,
but could result from central heating similar to
that described here.

X-ray holes formed by central heating 
are randomly disposed around the 
center of the flow, as in the Perseus cluster.
This is a particularly desirable feature since 
the X-ray bubbles need not be aligned with 
the accretion axis of the central black hole. 
Nevertheless, 
radio lobes often appear to be associated with X-ray holes.
This connection may be causal, but in some cases the 
relativistic electrons may readily 
flow into pre-existing holes; 
this may be occurring in the central southwestern hole in 
Perseus (Fig. 1 of Fabian 2001).

Finally, the azimuthally averaged 
hot gas density profile in these centrally heated models 
is flattened by the holes, 
providing an excellent fit to observed X-ray surface 
brightness profiles which are too centrally peaked otherwise.
To maintain this fit, AGN heating would need 
to reoccur every $\sim 10^8$ years. 
However, in the presence of central 
heating the gas temperature profile is less satisfactory, 
lacking the deep central minimum usually observed. 

We conclude that central heating in cooling flows 
naturally results in 
a relatively small number of 
randomly located X-ray bubbles similar to 
the irregularities observed.
In addition, 
these expanding plumes and bubbles are 
surrounded by considerably cooler gas, also consistent 
with recent {\it Chandra} observations. 
Gas in the cold rims was shock-heated 
but subsequently cooled by expansion.

\vskip.4in
Studies of the evolution of hot gas in elliptical galaxies
at UC Santa Cruz are supported by
NASA grant NAG 5-8049 and NSF grants  
AST-9802994 and AST-0098351 for which we are very grateful.
FB is supported in part by grants MURST-Cofin 00
and ASI-ARS99-74.

%\clearpage
\vskip.1in
\figcaption[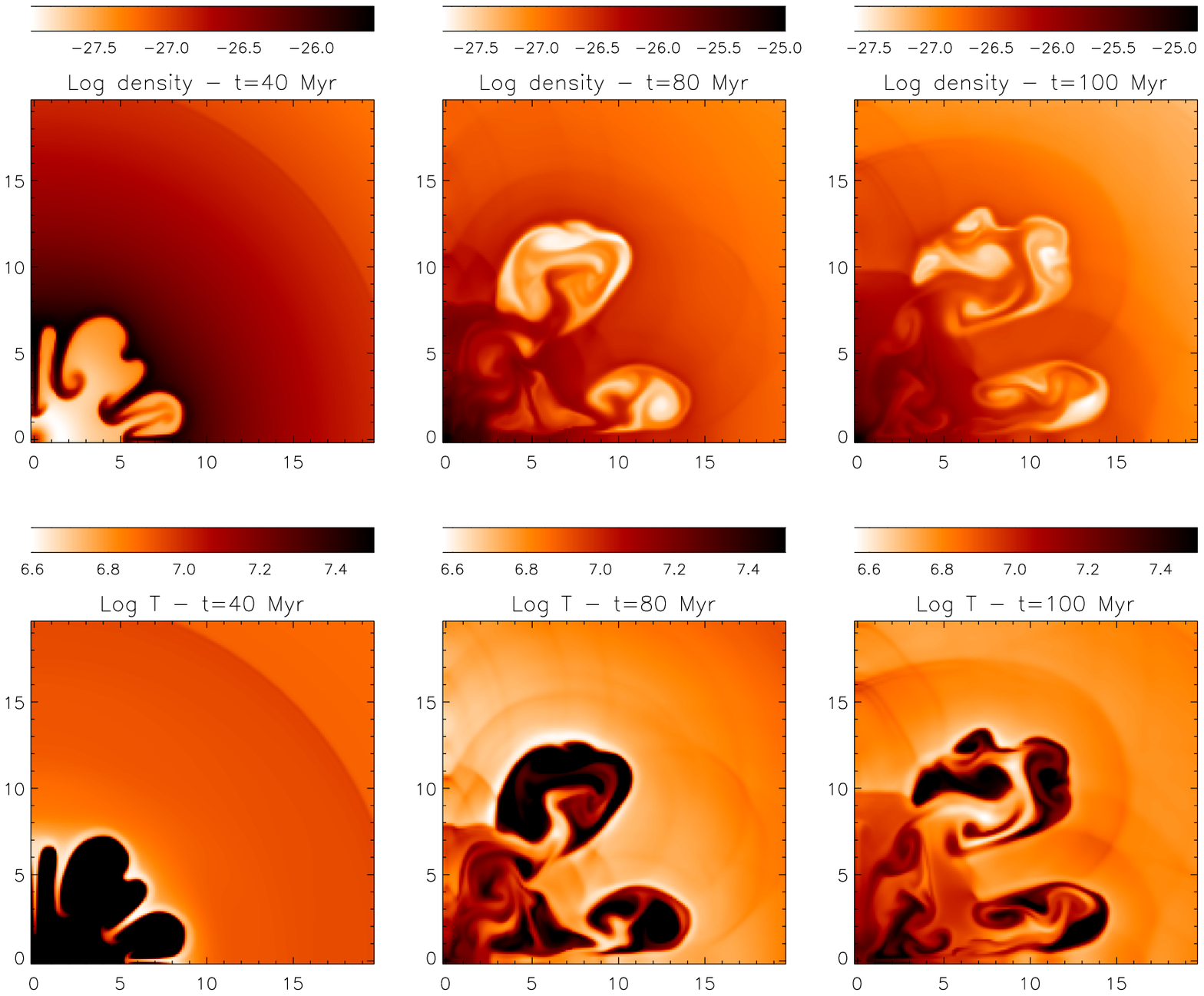]{
Evolution of centrally heated gas in a galactic scale cooling 
flow. 
The $z$ (vertical) and $R$ (horizontal) coordinates 
in each panel are labeled in kpc.
The upper and lower series of panels show respectively 
the gas density and temperature at three times:
$t_1 = 40 \times 10^6$,
$t_2 = 80 \times 10^6$, and $t_3 = 100 \times 10^6$ yrs 
from left to right.
Already at time $t_1$ 
the initial precursor shock has moved out to 
$\sim 20$ kpc and additional shocks are 
seen at time $t_3$. 
The hot plumes and bubbles are surrounded by 
relatively cold gas, particularly at early times.
This figure can be viewed in color at the electronic ApJ site.
\label{fig1}}

\vskip.1in
\figcaption[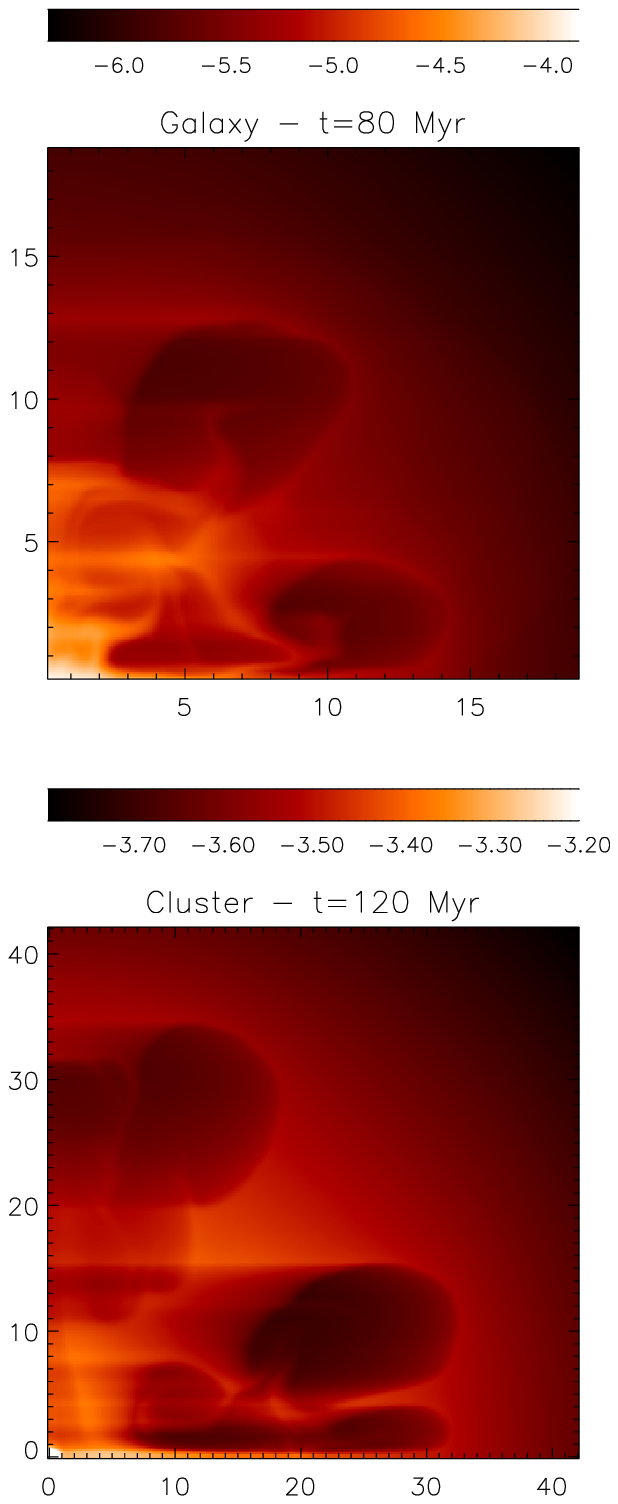]{
X-ray surface brightness distributions (0.5 - 2 keV) showing 
bubbles in two flows.
The $z$ (vertical) and $R$ (horizontal) coordinates
in each panel are labeled in kpc.
{\it Upper panel:} For the galactic 
scale cooling flow in Figure 1 
shown at time $t_2 = 80 \times 10^6$ yrs. 
{\it Lower panel:} 
For a rich cluster cooling flow at time
$t = 120 \times 10^6$ yrs
with heating parameters:
$L_h = 10^{45}$
ergs s$^{-1}$, $r_h = 2$ kpc and $t_h = 5 \times 10^7$ yrs.
This figure can be viewed in color at the electronic ApJ site.
\label{fig2}}

\vskip.1in
\figcaption[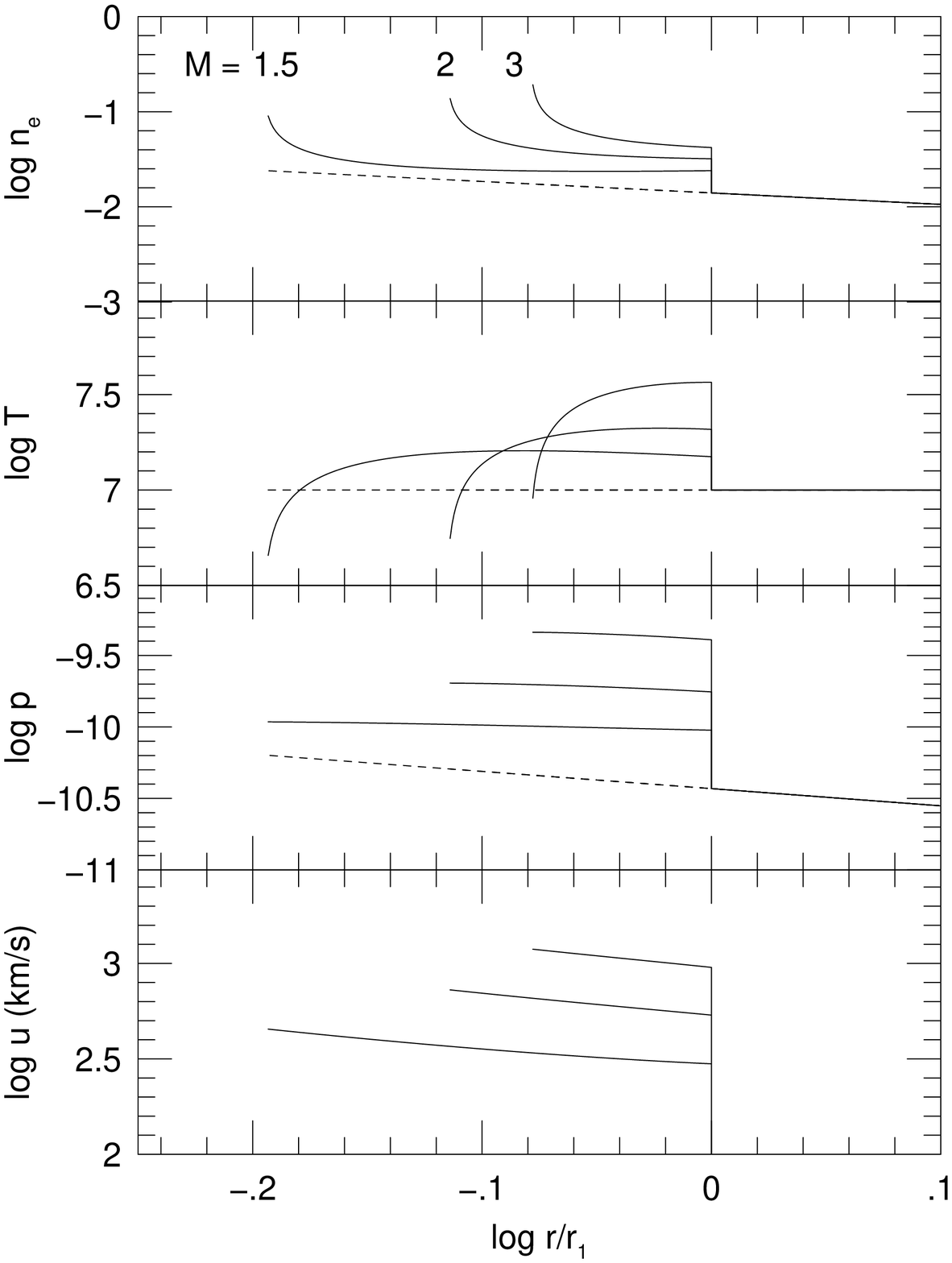]{
Self-similar flow of gas produced by a spherical piston 
expanding uniformly into an isothermal, 
fully ionized static gas with 
initial density 
$n_e = 0.0526 r^{-\alpha}$ cm$^{-3}$.
In descending order the panels show the radial 
dependence of the electron density, temperature,
pressure and velocity. 
The flow is shown at the (arbitrary) moment when the 
shock has arrived at $r_1 = 3$ kpc with 
Mach numbers 1.5, 2 and 3. 
The radial variation of the initially undisturbed gas
is shown with dashed lines. 
As suggested by the plots,
the temperature in the 
flowing gas near the piston (at the left of each  
plot) decreases to zero as the density becomes 
infinite.
\label{fig3}}


\begin{references}
\reference{} Blanton, E. L., Sarazin, C. L., McNamara, B. R. \&
Wise, M. W. 2001, ApJ 558, L15
\reference{} B\"ohringer, H. et al. 1993, MNRAS, 264, L25
%5044 Chandra images
\reference{} Brighenti, F. \& Mathews, W. G. 2002, 
ApJ (in press) (astro-ph/0203409)
%Three Virgo E galaxies:
\reference{} Brighenti, F. \& Mathews, W. G. 1997,
ApJ 486, L83
\reference{} Buote, D. A. et al. 2002, ApJ (in press)
(astro-ph/0000000)
\reference{} Caon, N., Macchetto, D. \& Pastoriza, M.
2000, ApJ 553, L125
\reference{} Chevalier, R. A. \& Imamura, J. N. 1983, ApJ 270, 554
%arc minute X-ray structure in 1275 with ROSAT:
\reference{} Churazov, E. et al. 2000, A\&A, 356, 788
%buoyant bubbles in M87:
\reference{} Churazov, E. et al. 2001, ApJ, 554, 261
\reference{} Courant, R. \& Friedrichs, K. O. 1948, Supersonic Flow
and Shock Waves (Interscience:London), ch. VIC
%Chandra images => feedback heating in Hydra A:
\reference{} David, L. P. et al. 2000, ApJ, 557, 546
%shocks moving from center or 4636:
David, L., Warmflash, A., Murray, S., \& Nulsen, P. E. J.
2002 (ApJ in press?) (astro-ph/0108114)
%Perseus/1275 holes:
\reference{} Fabian, A. C., Sanders, J. S., Ettori, S.,
Taylor, G. B., Allen, S. W., Crawford, C. S.,
Iwasawa, K., Johnstone, R. M. \& Ogle, P. M. 
2000, MNRAS, 318, L65
%holes in M84/NGC 4374:
%cold rims around all X-ray holes, a review:
\reference{} Fabian, A. C. 2002, (astro-ph/0201386)
%cold rims around holes:
\reference{} Fabian, A. C. et al. 2000, MNRAS, 318, L65
%more cold rims around holes in A1795:
\reference{} Fabian, A. C., Sanders, J. S., Ettori, S.,
Taylor, G. B., Allen, S. W., Crawford, C. S.,
Iwasaawa, K. \& Johnstone R. M. 
2001, MNRAS, 321, L33
%discussion of the absence of X-ray emission lines for cooling gas:
\reference{} Fabian, A. C., Mushotzky, R. F., Nulsen, P. E. J. \&
Peterson, J. R. 2001, 321, L20
\reference{} Finoguenov, A. \& Jones, C. 2001, ApJ, 547, L107
%holes and irregularities in NGC507:
\reference{} Forman, W., et al. 2001, in Lighthouses of the 
Universe Conference, August 2001 (astro-ph/0111526)
%(explosive?) holes in NGC 4636:
\reference{} Heinz, S., Choi, Y-Y, Reynolds, C.S., Begelman, M.C.,
2002, ApJ, 569, L79 
\reference{} Jones, C. et al. 2001, ApJL (in press) (astro-ph/0108114)
%Hydra A holes:
\reference{} Loewenstein, M., Mushotsky, R. F., Angelini, L., 
Arnaud, K. A. \& Quataert, E. 2000, ApJ 555, L21
\reference{} McNamara, B. 2000, ApJ, 534, L135
\reference{} Nulsen, P. E. J., David, L. P., McNamara, B. R., 
Jones, C., Forman, W. R. \& Wise, M. 2002, ApJ 568, 163
\reference{} Parker, E. N. 1961, ApJ 133, 1014
%theory of bubbles, feedback in 3D:
\reference{} Quilis, V., Bower, R. G. \& Balogh, M. L. 2001,
MNRAS, 328, 1091
%heating by jets
\reference{} Reynolds, C. S., Heinz, S. \& Begelman, M. C. 2001,
ApJ, 549, L179
%heating by jets
\reference{} Reynolds, C. S., Heinz, S. \& Begelman, M. C. 2002,
MNRAS, 332, 271
\reference{} Rogers, M. H. 1957, ApJ 125, 478
%theory of hot bubbles in cooling flows:
\reference{} Soker, N., Blanton, E. L. \& Sarazin C. L. 2002,
ApJ (submitted) (astro-ph/0201325)
\reference{} Stone, J. M. \& Norman, M. L. 1992, ApJS 80, 753
%correspondence between X-ray and optical line emission:
Trinchieri, G. \& Goudfrooij, P. 2002, A\&A (in press)
(astro-ph/0202416)
\end{references}
\end{document}